# Observation of valley Landau-Zener-Bloch oscillations and pseudospin imbalance in photonic graphene

Yong Sun[1,2], Daniel Leykam[3], Stephen Nenni[1], Daohong Song[4], Hong Chen[2], Y. D. Chong[5,6], and Zhigang Chen[1,4,*]

1 Department of Physics and Astronomy, San Francisco State University, San Francisco, California 94132
2 MOE Key Laboratory of Advanced Micro-Structured Materials, School of Physics Science and Engineering, Tongji University, Shanghai 200092, China
3 Center for Theoretical Physics of Complex Systems, Institute for Basic Science, Daejeon 34126, Republic of Korea
4 MOE Key Laboratory of Weak-Light Nonlinear Photonics, TEDA Applied Physics Institute and School of Physics, Nankai University, Tianjin 300457, China
5 Division of Physics and Applied Physics, School of Physical and Mathematical Sciences, Nanyang Technological University, Singapore 637371, Singapore
6 Centre for Disruptive Photonic Technologies., Nanyang Technological University, Singapore 637371, Singapore

Y.S. and D. L. contributed equally to this work.
*Correspondence to ZC (zhigang@sfsu.edu)

**Abstract** : We demonstrate inter-valley Bloch oscillation (BO) and Landau-Zener tunneling (LZT) in an optically-induced honeycomb lattice with a refractive index gradient. Unlike previously observed BO in a gapped square lattice, we show non-adiabatic beam dynamics that are highly sensitive to the direction of the index gradient and the choice of the Dirac cones. In particular, a symmetry-preserving potential leads to nearly perfect LZT and coherent BO between the inequivalent valleys, whereas a symmetry-breaking potential generates asymmetric scattering, imperfect LZT, and valley-sensitive generation of vortices mediated by a pseudospin imbalance. This clearly indicates that, near the Dirac points, the transverse gradient does not always act as a simple scalar force as commonly assumed, and the LZT probability is strongly affected by the sublattice symmetry as analyzed from an effective Landau-Zener Hamiltonian. Our results illustrate the anisotropic response of an otherwise isotropic Dirac platform to real-space potentials acting as strong driving fields, which may be useful for manipulation of pseudospin and valley degrees of freedom in graphene-like systems.

Bloch oscillation (BO) and Landau-Zener tunneling (LZT) have intrigued scientists for decades as fundamental phenomena predicted from quantum mechanics [1], and their direct observations have been achieved in a variety of systems ranging from semiconductor superlattices [2], cold atoms and Bose-Einstein condensates [3, 4], to photonic structures [5-8] and plasmonic waveguide arrays [9]. Recently, attention has been drawn to the idea that BO and LZT can be drastically altered in systems with Dirac-like dispersion, such as graphene and topological insulators [10-12]. For instance, by deforming a honeycomb lattice (HCL) for ultracold Fermi gases, creation and annihilation of Dirac points has been realized [10], allowing for probing the properties of BO and LZT as the system undergoes a topological transition [11, 12]. In particular, artificial Dirac systems [13] such as "photonic graphene" - a HCL of evanescently coupled optical waveguides [14, 15], and "synthetic solids" of trapped ultracold atoms in crystals of light [10, 16, 17], have provided a tunable platform as a quantum simulator for many fundamental phenomena, including demonstration of photonic Floquet topological insulators [15] and the measurement of Bloch band topology and Berry curvature [17].

While achievements have been made in observation of spatial shifts of wavepackets undergoing adiabatic BO using cold atoms in optical lattices [3], the non-adiabatic phenomenon of LZT involving Dirac points under strong driving field is still poorly understood. Thus far, the systems in which BO and LZT have been most extensively studied are one-dimensional gapped periodic lattices. In such systems, the dynamics of BO and LZT are highly sensitive to the gap size: for large gaps adiabatic transport within a single energy band results in periodic BO, but smaller gaps comparable to the driving force lead to interband LZT that can break the periodicity of the dynamics and induce net transport [18]. In the limit of a vanishing gap, period-doubled BO is expected to be restored due to perfect LZT [19]. This behavior becomes more complex in two-dimensional (2D) systems, where the Bloch oscillation trajectories and LZT strength may become sensitive to the direction of the applied field [20]. In optics, BO and LZT in the 2D domain were previously demonstrated only with square photonic lattices [8], but the physics in graphene-type HCL [14] is fundamentally different: In square lattices, LZT occurs through gapped Bloch bands with no Dirac points involved, whereas in HCL the band gap vanishes at the Dirac points and the LZT probability is sensitive to their relative chirality as well as the direction of driving force. Hence, the interplay of BO (between valleys) and LZT (near the Dirac points) is expected to bring about wavepacket dynamics mediated by pseudospin [21] and valley degrees of freedom qualitatively distinct from the behavior in square lattices. To our knowledge, BO through Dirac points [22] and the role played by pseudospin have so far never been observed.

In this Letter, we demonstrate inter-valley BO and LZT at Dirac points using optically-induced photonic graphene. We observe non-adiabatic wavepacket dynamics that depend anisotropically on the direction of an applied potential gradient. For one choice of the gradient (one that does not break the sublattice symmetry), we observe persistent and symmetric BO through two *inequivalent* valleys and perfect LZT, as expected for a massless 2D Dirac Hamiltonian driven by a scalar potential gradient. In contrast, for another choice of the gradient (one that breaks the sublattice symmetry), we observe damped BO between two *equivalent* valleys due to asymmetric scattering and imperfect LZT and, counterintuitively, the tunneling probability decreases as the driving field strength is increased. This latter scenario suggests that, near the Dirac points, the potential gradient does not always act as a simple scalar force as commonly assumed. Our theoretical analysis of the LZT probability based on an effective Landau-Zener Hamiltonian along with calculations of BO dynamics to long distances shows unambiguously the influence of the sublattice (pseudospin) symmetry breaking. Moreover, the broken symmetry leads to a pseudospin imbalance, observable in the form of vortices with valley-dependent topological charges. These results demonstrate clearly that, as shown in many occasions in modern physics, dispersion does not necessarily determine everything by itself, and equally important is the structure of the eigenstates, unveiling how isotropic Dirac cones can exhibit an anisotropic response to strong driving fields.

Our experimental setup is shown in Fig. 1(a), where the HCL is optically induced in a biased photorefractive crystal (SBN:60) by six interfering ordinarily-polarized beams obtained through an amplitude mask with appropriate phase modulation [14, 21]. The HCL intensity distribution (with the orientation of one principal axis of the lattice along the *x*-direction) is described by $I_b = \frac{4A}{27}|\sin(\sqrt{3}\pi x/d - \pi y/d) + \sin(\sqrt{3}\pi x/d + \pi y/d) - \sin(2\pi y/d)|^2$, where $A$ is the peak intensity, $x$ and $y$ are the transverse coordinates, $d = \sqrt{3}a/2$, and $a = 31\,\mu\text{m}$ is the lattice period. Such a periodic light pattern remains stationary along the whole length of the 2-cm crystal as shown in Fig. 1(b), which induces a waveguide array (HCL) discretizing the diffraction of an extraordinarily polarized probe beam. Figures 1(c) and 1(d) are the measured output discrete diffraction pattern and corresponding Brillouin zone (BZ) spectrum [23] when probed with a focused partially coherent beam, demonstrating the fidelity of the photonic HCL used in our experiments. In addition, a transverse refractive-index gradient is induced by laterally illuminating the crystal with white light, whose intensity is modulated by inserting halfway a razor blade. This white-light illumination is well-approximated by $I_m = B[1 + \tanh(y/\eta)]$, where $B$ is the background illumination, and the parameter $\eta$ determines the extent of the induced index ramp [8]. The total induced refractive-index change can be written as $\Delta n = \gamma(I_m + I_b)/(1 + I_m + I_b)$, where $\gamma$ is the normalized nonlinear coefficient that can be tuned by varying the bias field.

The propagation of the probe beam is described by the paraxial Schrodinger-type equation for the normalized electric field envelope $\psi(x, y, z)$ [14],

$$i\frac{\partial \psi(x,y,z)}{\partial z} = -\frac{\lambda}{4\pi n_0}\nabla_\perp^2 \psi(x,y,z) - \frac{2\pi}{\lambda}\Delta n(x,y)\psi(x,y,z), \tag{1}$$

where $z$ is the propagation distance, $\lambda = 488\,\text{nm}$ is the laser wavelength in vacuum, and $n_0=2.36$ is the unperturbed refractive index of the SBN crystal. For the numerical simulations of Eq. (1), we employ a standard split-step beam propagation method with parameters chosen according to our experimental conditions: lattice intensity $A=0.075$, background illumination $B=2A$, and normalized nonlinear coefficient $\gamma = 2.2\times 10^{-3}$.

First, we consider Bloch oscillations induced by an index gradient $\eta = 250\,\mu\text{m}$ imposed parallel to the direction of $x$-axis, along which the top and bottom zigzag edges are oriented. The resulting BO period is 3.3cm, as estimated from the accumulation of transverse momentum along the BO direction. As shown in Fig. 1(b), in this case A and B sites within each unit cell experience the same index potential and, consequently, the sublattice symmetry is preserved. In the experiment we cannot directly monitor the beam evolution throughout the crystal, but we can emulate its behavior by observing the output profile in momentum space for various input tilts of the probe beam. The output Fourier spectra of the probe beam shown in Figs. 2(a-e) along with corresponding numerical simulations in Figs. 2(f-j) give a fully conclusive picture of the beam evolution during one BO through the K and K' valleys, as illustrated by horizontal dashed lines in Fig. 1(d).

As shown in Fig. 2(a), the initial excitation of the first band displays strong resonant scattering when it reaches one of the K valleys, symmetrically populating the other two equivalent K valleys. In principle, each scattered spectral component contains two parts: one belonging to the second Bloch band due to LZT, and the other belonging to a Bragg-reflected first band component that remains in the first BZ. Experimentally, the 2$^{nd}$-order BZs are associated with but not equivalent to the second Bloch band, which complicates the determination of the tunneling rate through a particular Dirac point. Theoretically, we know that for a sufficiently weak gradient, significant tunneling occurs only in the vicinity of the Dirac point, which is described by a Landau-Zener Hamiltonian of the form [22]

$$H(z) \approx \Delta \hat{\sigma}_z - \frac{\sqrt{3}}{2} azEC\hat{\sigma}_x, \qquad (2)$$

where $\Delta$ is the index contrast between the two sublattices, $\hat{\sigma}_j$ are Pauli matrices, $E$ is the transverse index gradient, $C$ is the effective hopping strength between neighboring lattice sites, and we have used a $z$-dependent momentum gauge $\mathbf{p}(z) = \mathbf{p}_0 + \mathbf{E}z$ to represent the linear index gradient as a constant force acting on the wavepacket. The corresponding LZT probability is $P_{LZ} = \exp\left(-2\pi\Delta^2 / \sqrt{3}|\mathbf{E}|Ca\right)$ [24, 25]. Because the sublattice symmetry is unbroken under an $x$-index-gradient, the effective mass at the Dirac point vanishes, $\Delta = 0$, and thus nearly perfect tunneling probability to the second band is expected regardless of the strength of the index gradient. When the input tilt angle is further increased in Figs. 2(b-e), the two scattered components accelerate towards the inequivalent K' valleys. Right after passing the K' valleys, they are scattered predominantly back to the 1$^{st}$ BZ to complete one Bloch oscillation cycle. Thus, the energy of the probe beam mostly returns to the first band, consistent with the prediction of perfect LZT at each Dirac point and coherent Bloch oscillations within the two lowest bands.

Next, we consider an index gradient $\eta = 400\mu$m parallel to the *y*-axis, i.e. parallel to one of the reciprocal lattice vectors, which induces BO between two equivalent K valleys with a period of 3.1cm. As seen from Figs. 1(b, d), in this case the gradient breaks the sublattice symmetry as A and B sites within each unit cell experience a different index potential. This in turn lifts the degeneracy of pseudospin states [21] but preserves the valley degree of freedom. Typical results are presented in Fig. 3. The probe beam is accelerated through the first BZ, and then scattered at the top K valley as shown in Figs. 3(a-c). Clearly, the resonant scattering to the other two equivalent K points is highly asymmetric. After completing one BO cycle in Fig. 3(e), the beam is split between the first and second BZs, indicating strong but *imperfect* LZT. To explain this, we assume that the effective Hamiltonian has a nonzero mass term that is proportional to the applied potential difference: $\Delta = \boldsymbol{E} \cdot \boldsymbol{\delta}/2$, where $\boldsymbol{\delta} = a/\sqrt{3}\,(0,1)$ is the displacement between the A and B sublattices. When $\Delta$ vanishes as in the case with an *x*-gradient, a perfect LZT occurs, but $\Delta$ is now nonzero with a *y*-gradient and it is proportional to the index gradient. In this latter case, $P_{LZ} = \exp\left(-\pi a|\boldsymbol{E}|/6\sqrt{3}C\right)$ decreases *monotonically* with the driving potential. Thus, any nonzero gradient splits the beam between the two bands, with isotropic near-perfect LZT occurring only in the weak field limit $a|\boldsymbol{E}| \ll 1$. This anomalous tunneling probability is a unique characteristic of Dirac points, dependent on the mass term being proportional to the applied force, and it is not observable in conventional gapped bands such as in 2D square lattices [8]. Detailed analysis of the anisotropic LZT probability at the Dirac points based on the effective Landau-Zener Hamiltonians (supported by the tight-binding calculations of the wavepacket dynamics) can be found in the Supplementary Information [25].

To verify that the Dirac points are responsible for the strong LZT observed for both index gradients, we repeated the experiment under the same conditions, except for using a probe beam trajectory along the Γ and M points to avoid the valleys (see Fig. 1(d)) for direct comparison. In this case, as shown in the bottom row of Fig. 3, negligible LZT is observed, i.e., $P_{LZ} \to 0$, and almost complete restoration of the input beam after one BO cycle. While these periodic dynamics resemble the BO shown in Fig. 2, in this case the wavepacket always remains in the first Bloch band.

In order to provide a quantitative comparison of BO amplitude and its dynamical behavior, we plotted in Fig. 4 the calculated beam center of mass from simulation of beam propagation to a much longer propagation distance (than the experimentally accessible length), showing persistent coherent BO for the *x*-gradient but severely damped BO for the *y*-gradient. As see clearly in Fig. 4(a), the *x* component of the center of mass oscillates with a regular period of 3.3cm and persists for long distances exceeding 12cm, indicating coherent BOs and a low LZT probability to higher bands. However, while for the *y*-gradient the BO quickly washes out (see Fig. 4(b)) due to splitting of the beam between the two bands, despite the fact that a weaker gradient is used. In this latter case, imperfect LZT splits the beam into two each time the Dirac point is crossed, and the split components quickly lose their mutual coherence, washing out the spatial oscillations during subsequent propagation. It should be pointed out that, different from the case of *y*-gradients, the BO becomes strictly periodic in the tight-binding model for the case of weaker *x*-gradients when the radiation losses/tunneling to higher bands is prevented, as detailed in the Supplementary Material [25].

The gradient-induced symmetry breaking of the two sublattices does not only affect the interband tunneling probability, but it also lifts the degeneracy of the pseudospin basis states. Away from the Dirac points, this effect is much weaker than the inter-site coupling, so the instantaneous eigenstates of Eq. (2) are simply the Bloch waves of the undriven lattice, which excite the two sublattices equally and have vanishing pseudospin, i.e. $\langle \sigma_z \rangle \approx 0$. However, at momenta close to the Dirac points, the symmetry breaking term $\Delta$ dominates and the instantaneous eigenstates reside on separate sublattices, i.e. $\langle \sigma_z \rangle = \pm 1$. Therefore, during the BO, an equal excitation of the two sublattices (i.e. with vanishing pseudospin) is converted into one exhibiting a pseudospin imbalance, leading to the generation of pseudospin angular momentum [21]. The imbalance is maximum in the vicinity of the Dirac points, where the pseudospin eigenstates consist of the superposition of the three equivalent $K_{1,2,3}$ points with a vortex phase winding, $(K_1, K_2, K_3) = (1, e^{\pm i2\pi/3}, e^{\mp i2\pi/3})$. We can measure this imbalance-induced vorticity by reducing the size of the input probe beams in real space (i.e., expanding them in Fourier space), so that the tails of the probe beams can overlap and undergo interference.

Typical experimental results from measuring such a vortex phase are shown in Fig. 5, where a relatively narrow beam is used as the probe. In the absence of an index gradient [Fig. 5(a)], there is no pseudospin generation since both sublattices are equally excited [21], and hence no vorticity is observed at the output. Similarly, even with the *x*-gradient, the sublattice symmetry remains unbroken so no net vorticity is observed at the output [Fig. 5(b)]. In this case, a vortex-antivortex pair is noticeable in the first BZ, indicating that the *x*-gradient (acting as a synthetic field) gives the pseusdospin states a relative phase [26]. In contrast, when the *y*-gradient is applied, the broken sublattice symmetry induces a pseudospin imbalance and net vorticity. Furthermore, switching to an inequivalent valley reverses the topological charge [Figs. 5(c,d)], as required by time reversal symmetry. In the two right panels of Fig. 5, such nontrivial vortex phase under different excitation conditions is plotted from numerical calculation. Thus, the vortex generation during the valley-conserving BO provides a measure of the chirality of the Dirac points. We stress that the pseudospin imbalance and corresponding topological charge observed here is determined by the LZT dynamics, which are sensitive to parameters such as the index gradient, lattice potential depth, and propagation distance. Near the Dirac points, the populations of the two bands oscillate rapidly [24]. Nevertheless, for a given set of parameters, reversing the valley index (i.e., from K to K' or vice versa) must reverse the topological charge, as observed in our experiment [Figs. 5(c,d)]. However, reversing the direction of index gradient should not affect the tunneling rate. Thus, for a valley-conserving index gradient, the LZT is sensitive to the choice of valleys through the generation of a sublattice pseudospin imbalance.

Our results show that there is an important difference between the real electrons described by the Dirac equation and the electrons in graphene (with the sublattice pseudospin, for example), and that special attention should be paid to such counter-intuitive effects. As a typical example, a mechanism similar to that presented in this work can also affect the dynamics of Klein tunneling in graphene [27], if a potential gradient is introduced to locally induce mass in the dispersion by breaking the sublattice symmetry.

In conclusion, we have observed valley-dependent BO and LZT in photonic graphene driven by an index gradient. The interplay between the applied index gradient, sublattice symmetry breaking, and Bragg scattering can generate beam dynamics that are sensitive to the choice of the Dirac cones. Our observations of asymmetric scattering, imperfect LZT, and valley-sensitive generation of vortices mediated by pseudospin imbalance reveal the anisotropic response of Dirac points in an otherwise isotropic platform of HCL to strong driving fields, which may be used to control valley and pseudospin degrees of freedom in graphene-like systems. Our results may also provide insight to recent relevant studies involving BO in parity-time-symmetric structures [28, 29], flat-band systems [30, 31], as well as spin-orbit coupled systems [32].

We thank Y. Kivshar and D. Neshev for discussions, and referees for their insightful comments. This work is supported by ARO and NSF, and by the National Key R&D Program of China (2017YFA0303800, 2016YFA0301101), the Chinese National Science Foundation (11674247, 91750204, 18ZR1442900), the Singapore National Research Foundation (NRFF2012-02) and MOE Academic Research Fund Tier 2 grant (MOE2015-T2-2-008), and the Institute for Basic Science in Korea (YSF-R024-Y1).

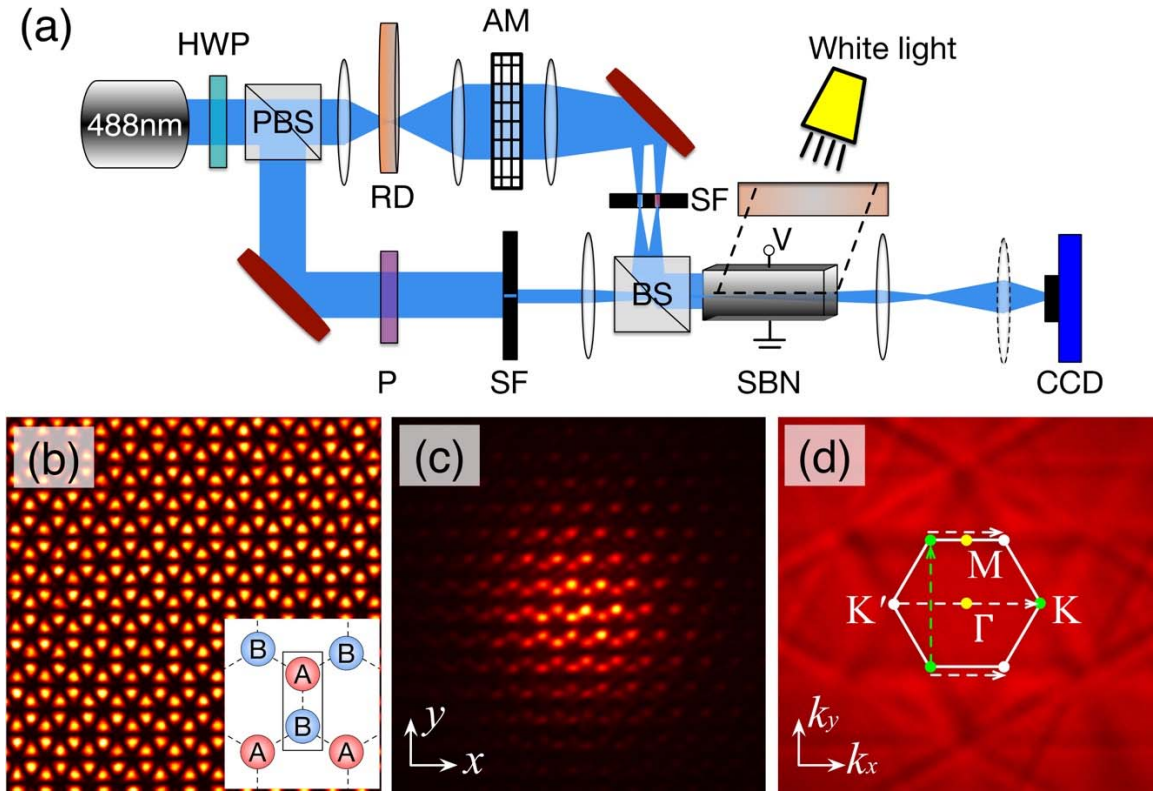

FIG. 1: (a) Schematic of experimental setup. Upper path is for optical induction of photonic graphene, and bottom path is for probing through the lattice. Refractive index ramp is induced by nonuniform white light illumination. HWP: half-wave plate; PBS: polarizing beam splitter; RD: rotating diffuser; AM: amplitude mask; P: polarizer; SF: spatial filter; SBN: strontium barium niobate. (b) Output HCL intensity pattern. Inset: illustration of A/B sublattice structure. (c) Discrete diffraction of a Gaussian beam exiting the lattice. (d) Measured spectrum superimposed with schematic drawing of the 1st BZ, where symmetry points are marked by dots and paths for BO are depicted by dashed lines.

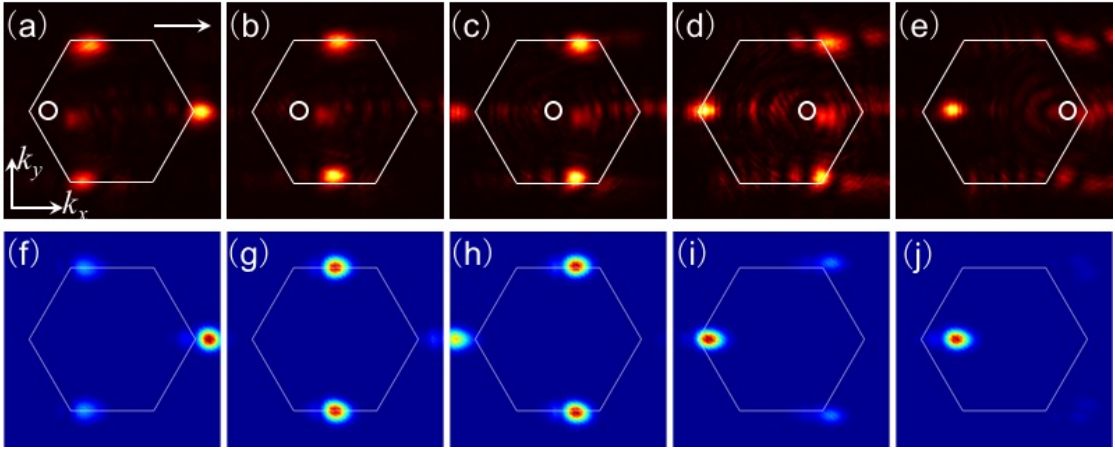

FIG. 2: Symmetric BO under an *x*-index-gradient. (a-e) Measured Fourier spectra at the lattice output for different input tilts of the probe beam (marked by the white circles). The arrow in (a) indicates the direction of the gradient. In (a), when the probe beam crosses the Dirac K point, there is symmetric scattering to the other two equivalent K points. In (d, e), the two scattered components cross the K′ points, and then predominantly return to the first BZ. (f-j) Corresponding results from numerical simulation. The path of this BO is illustrated by horizontal dashed lines in Fig. 1(d).

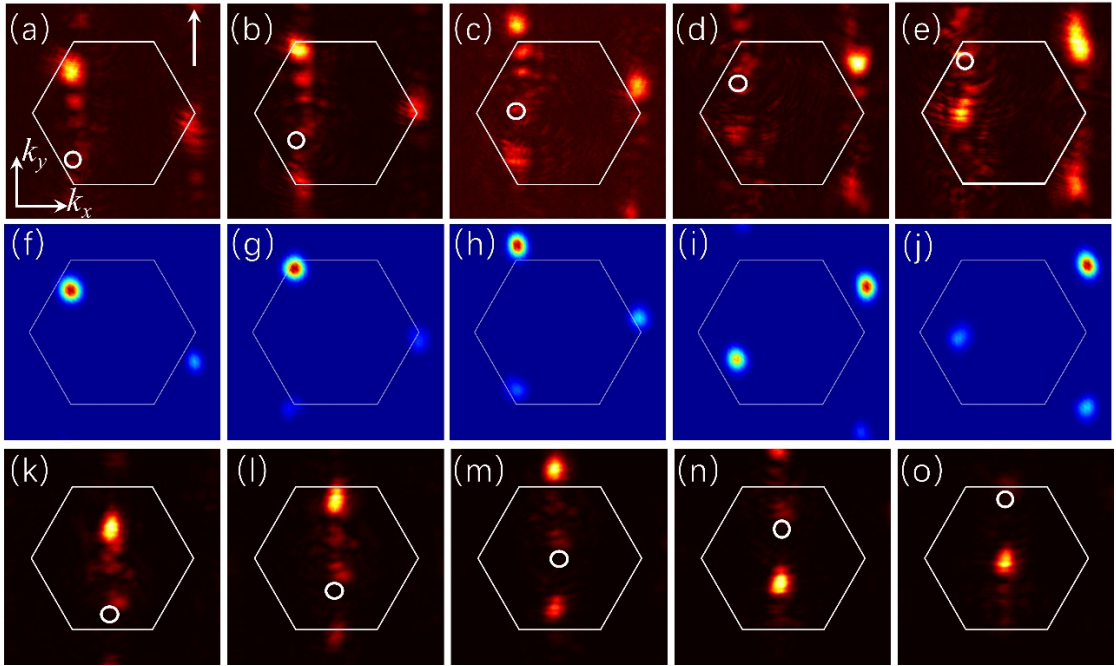

FIG. 3: Top two rows: Asymmetric BO under a *y*-index-gradient. (a-e) Same as in Figs. 2(a-e) except that the index gradient is now in *y*-direction. In (b-e), when the probe beam crosses the Dirac K point, the scattering to the other two equivalent K points is highly asymmetric, leading to imperfect tunneling into the 2nd band. (f-j) Corresponding numerical results. The path of this BO is illustrated by a vertical dashed line in Fig. 1(d). The bottom row shows BO through $\Gamma$ and M symmetry points (no valley involved), corresponding to quasi-adiabatic transport primarily in the 1st band.

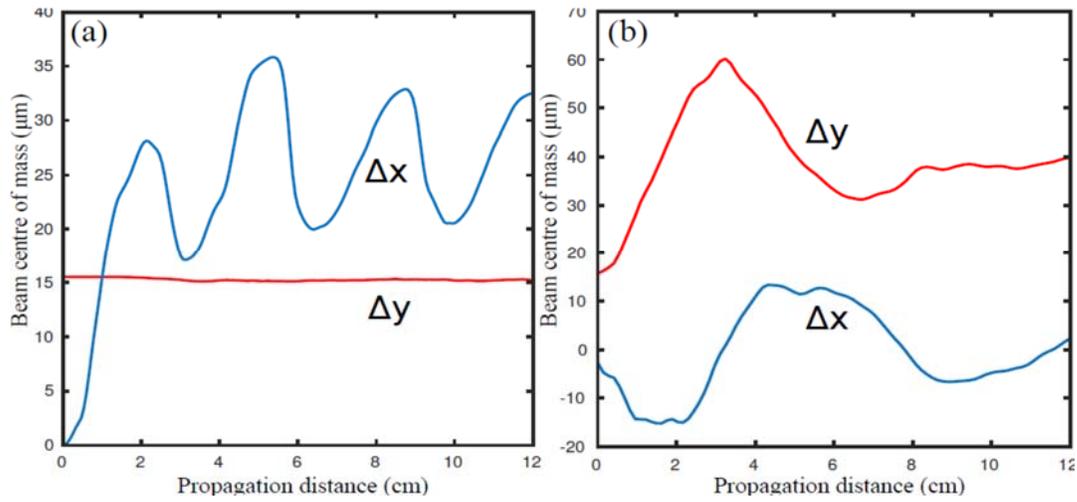

FIG. 4: Calculated real-space displacement of the beam's "center of mass" for propagation beyond experimentally accessible distance for direction comparison of the BO dynamics. (a) Coherent BO under an $x$-index-gradient, displaying its persistence along the $x$-direction. (b) Asymmetric BO under a $y$-index-gradient, showing faded BO due to imperfect LZT and loss of coherence of split beams upon traversing the Dirac point.

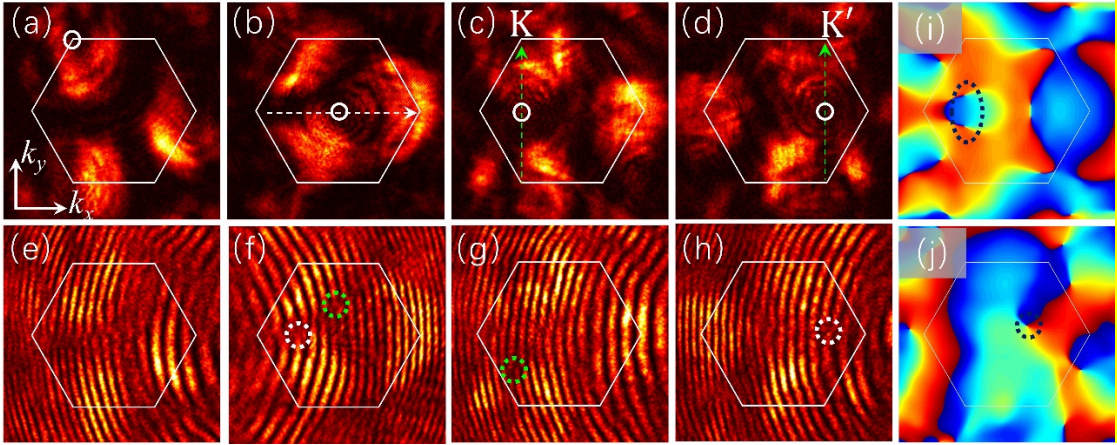

FIG. 5: Valley interference after BO probed with a broad beam. (a-d) Measured output intensity under (a) no gradient, (b) horizontal gradient, (c) vertical gradient for BO along K point, and (d) vertical gradient for BO along K′ point. White circle indicates the input beam position. (e-h) Measured interferograms showing fringe forks (vortex positions) in the first BZ marked by dashed circles. (i, j) Phase structures obtained from numerical simulations corresponding to (f, g).